\documentclass[aps,prb,reprint,showpacs,preprintnumbers,amsmath,amssymb]{revtex4-1}
\usepackage{graphicx}

\begin{document}

\frenchspacing

\title[Chalcogenide nanoclusters]{X-ray total scattering study of regular and magic-size nanoclusters of cadmium sulphide}
\author{Lei Tan,  Alston J Misquitta, Andrei Sapelkin}
\affiliation{School of Physics and Astronomy, Queen Mary University of London, Mile End Road, London, E1 4NS,UK}
\author{Le Fang}
\affiliation{School of Biological and Chemical Sciences, Queen Mary University of London, Mile End Road, London, E1 4NS, UK}
\author{Rory M Wilson}
\affiliation{School of Engineering and Materials Science, Queen Mary University of London, Mile End Road, London, E1 4NS, UK}
\author{Baowei Zhang, Tingting Zhu, Frank S Riehle, Shuo Han, Kui Yu}
\affiliation{Institute of Atomic and Molecular Physics, Sichuan University, Chengdu, 610065, P. R. China}
\author{Martin T Dove}
\email[Corresponding author: ]{martin.dove@qmul.ac.uk}
\affiliation{School of Physics and Astronomy, Queen Mary University of London, Mile End Road, London, E1 4NS, UK, and School of Physics, Sichuan University, Chengdu, 610065, P. R. China}

\begin{abstract}
Four kinds of magic-size CdS clusters and two different regular CdS quantum dots have been studied by x-ray total scattering technique and pair distribution function method. 
Results for the regular CdS quantum dots could be modelled as a mixed phase of atomic structures based on the two bulk crystalline phases, which is interpreted as representing the effects of stacking disorder. However, the results for the magic-size clusters were significantly different. On one hand, the short-range features in the pair distribution function reflect the bulk, indicating that these structures are based on the same tetrahedral coordination found in the bulk phases (and therefore excluding new types of structures such as cage-like arrangements of atoms). But on the other hand, the longer-range atomic structure clearly does not reflect the layer structures found in the bulk and the regular quantum dots. 
 We compare the effects of two ligands, phenylacetic acid and oleic acid, showing that in two cases the ligand has little effect on the atomic structure of the magic-size nanocluster and in another it has a significant effect.

\end{abstract}

\maketitle

\section{introduction}
Metal chalcogenide semiconductor nanoclusters, which form as quantum dots (QDs)\cite{Yin2005,Qin2014,Dolai2015}, have been explored for applications in biomedical labelling \cite{Zhang2016}, solar cells \cite{Dolai2015}, light emitting diodes \cite{Dolai2015} etc.
 Typically QDs have their surfaces coated by ligands to passivate the surfaces \cite{Corsini2017}, shape the QDs\cite{Alivisatos1996}, and protect the QDs from aggregation and corrosion\cite{Tan2012}. Moreover, these coatings can tune the optical properties \cite{Chen2011,Tohgha2013}.
 
The quantum dots are found either as regular quantum dots (RQDs) or magic-size clusters \cite{Yu2012} (MSCs), which are normally in the size range  up to around 30 \AA \cite{Dolai2015}. 
MSCs contain a well-defined number of atoms \cite{Kudera2007} and thus a single size. The high stability of MSCs enables them to form so-called ``cluster assembled materials'', providing a promising approach for the design of nano-devices \cite{Zhang2015}. 

The strong quantum confinement of electrons in quantum dots means they have size-dependent electrical, optical, magnetic and catalytic phenomena\cite{Gary2016,Tohgha2013}, which cannot be realised by their bulk counterparts \cite{Kim2014}. However, the effect of detailed atomic structure has been less well understood, and some important aspects for both CdS RQDs and MSCs are explored in this paper.

Our method of choice is x-ray total scattering method\cite{Masadeh2007,Billinge2007}, which directly gives us real-space structural information through the pair distribution function (PDF), which is closely related to the histogram of interatomic distances\cite{Billinge2007}. Standard methods such as transmission electron diffraction can give some information on size and shape, but this is tempered by the fact that there can be too low a contrast between the nanocluster cores and ligands \cite{Kim2014,Gary2015}. In favourable circumstances it is possible to grow single crystals from self-aggregation of MSCs\cite{Gary2016}, but we do not have this good fortune for CdS. Thus total scattering is our preferred method here.

In this paper we report a number of new insights into the atomic-scale structure of the cores of CdS RQDs and MSCs. Comparison of PDFs from the bulk sample and RQDs, combined with structure fitting, gives a diameter for the RQDs  and shows that the atomic structures of the RQDs are closely related to the layer structures of the bulk. 
On the other hand, our PDF measurements show that the atomic structures of the MSCs with two different ligands contain the same local tetrahedral coordination as in the bulk, but that neighbouring tetrahedra are connected without the layer arrangements of the bulk. An atomic model based on a recent crystal structure refinement for a self-assembled arrangement of InP MSC \cite{Gary2016} gives a model for one CdS MSC that reproduces well the data presented here. We will show that two other MSCs similarly have the same local structure but different longer-scale arrangements of atoms. The work presented here definitively excludes the possibilities that the MSCs have cage-like atomic structures. In the following sections we first discuss the methods we have used, then consider separately the new insights into the atomic-scale structures of the RQDs and MSCs.

\section{Methods}
\subsection{Sample synthesis and characterisation}

\subsubsection{Magic-size clusters}

The single-size CdS MSCs with phenylacetic acid (PA) ligands were prepared using two-step approaches. The first step involved heating Cd and S precursors at a relatively high temperature, and the second step involved cooling the resulting mixture to a relatively low temperature and dispersing it into a solvent such as cyclohexane or toluene\cite{Zhu2017, Zhang:inpress, Liu2017, Zhu2018}. The single-size CdS MSCs with oleic acid (OA) ligands were synthesised by a one-pot method \cite{Nevers2017, Nevers2018}. Details are given in the electronic supplementary information(ESI). 

Each sample of CdS MSCs was characterised on the basis of their ultraviolet visible (UV-vis) absorption spectrum, and in this paper we follow standard practice and label different NCs on the basis of the wavelength of the peak in the absorption spectrum, with units of nm.
The UV-Vis spectra of the four CdS MSC are given in Figure S1 of the ESI. These show sharp absorption peaks at 311 nm and 322 nm as seen in previous work.

The surface ligands were characterised by Fourier transform infrared spectroscopy (FTIR) as described in the ESI. We were able to confirm the presence of the PA and OA ligands in the NC sample.


\subsubsection{Regular quantum dots}

Two methods were used for the preparation of RQDs. The first, which led to a sample with UV-vis absorption peak at 355 nm (Figure S1 of the ESI), was prepared following the same method to generate the PA MSCs, but the heating in the second stage was at the higher temperature of 110~$^{\circ}$C.

In the second method  (author FSR) the RQDs were synthesised at a relatively low temperature from organo-metallic precursors in a weakly-coordinating solvent at ambient condition. Details about the synthetic method will be reported elsewhere. A new multi-step purification protocol for the effective removal of organic compounds was developed allowing the `interference-free' characterisation of the inorganic nano-sized CdS core material. Details about the purification process are reserved due to a patent application being prepared.  The RQD produced this way had an absorption peak at 439 nm (Figure S1).



\subsection{Total scattering measurements and extraction of the pair distribution function}
%

The x-ray total scattering measurements were performed on a PANalytical Empyrean diffractometer with a silver anode X-ray tube within the QMUL laboratory, with measurement times of 20 hours for scanning of scattering angle to $148^\circ$. Samples were loaded into kapton tubes of diameter 1 mm. In addition to performing the measurement of the sample, separate measurements of the empty kapton tube and of the empty instrument were performed for data correction purposes. The sets of measurements were combined and converted using the GudrunX program \cite{Soper2011}, which generates an overall scattering function $S(Q)$, where the scattering vector $Q = 4 \pi \sin \theta / \lambda$, with $\theta$ equal to half the scattering angle, and $\lambda = 0.5609$~\AA\ is the x-ray wavelength.

The conversion of the scattering data to the PDF was performed based on the following equations. Writing $\rho$ as the overall number of atoms per unit volume (number density), and $c_n \rho$ as the density of atoms as type $n$, we can express the number of atoms of type $n$ lying within a spherical shell of thickness $\mathrm{d}r$ at a distance $r$ from an atom of type $m$ as $4\pi r^2 \mathrm{d}r \times c_n \rho \times g_{mn}(r)$, where the partial PDF $g_{mn}(r)$ represents the difference of the distribution from random. The partial PDFs $g_{mn}(r)$ are combined as
\begin{equation}
D(r) = 4\pi\rho r \sum_{m,n}c_m c_n f_m f_n \left( g_{mn}(r) - 1 \right) 
\end{equation}
where $f_m$ is the atomic scattering factor.

The Fourier transform of this function gives the experimentally-derived scattering function:
\begin{equation}
Qi(Q) =  \frac{2}{\pi} \int_0^\infty D(r) \sin(Qr) \, \mathrm{d}r
\end{equation}
where $i(Q) = S(Q) - S(Q\rightarrow \infty) - S_0$, with $S(Q)$ being the measured function after corrections for sources of additional scattering and attenuation, and after normalisation. The component $S(Q\rightarrow \infty)$ arises from the self-scattering term proportional to $\sum_m c_m f_m^2$, and $S_0$ is scattering around $Q \sim 0$. In turn, the function $Qi(Q)$ has the reverse transformation:
\begin{equation}
D(r) =  \frac{2}{\pi} \int_0^\infty Qi(Q) \sin(Qr) \, \mathrm{d}r
\end{equation}
which is the key equation for converting the scattering data to the PDF.

Whilst there are many formulations of the PDF \cite{Keen2001}, the function $D(r)$ is our function of choice for analysis, because it is the fundamental Fourier transform of the data.
GudrunX \cite{Soper2011} was used to convert the scattering function $i(Q)$ to $D(r)$, exploiting a number of features that can be used to remove the inevitable effects of noise albeit at the cost of a small broadening of the resultant PDF. 

\subsection{Lattice simulations}
To support the experimental analysis we have performed calculations of the PDF using a special module within the General Utility Lattice Program (GULP) \cite{GULP_paper}, as described in reference \onlinecite{Cope2007}. The key point of this module is to calculate the widths of the Gaussian functions that describe the individual peaks in the PDF based on the underlying phonon frequencies and eigenvectors and a given temperature.

For these calculations, we used the simple force field model of Fan et al. \cite{Fan2014}, who used interaction potential functions of the form
\begin{equation}
U_{ij} = \frac {q_i q_j } {4 \pi \epsilon_0 r_{ij}}  + A \exp(-r_{ij} / \rho)  -\frac{C } {r_{ij} ^6}
\end{equation}
The first term describes the long-range Coulomb interactions. The second and the third terms are the Buckingham potential describing the repulsive and dispersion interactions. Parameter values of  $A$ , $\rho$ and $C$  for Cd--Cd, Cd--S and S--S pairs were derived by Fan et al. \cite{Fan2014}.

To simulate the final size of a nanoparticle, it is assumed that the nanoparticle has spherical shape and that the PDF of the bulk is modulated by the standard spherical envelop function \cite{Gamez-Mendoza2017,Gilbert2008}:
\begin{equation}
f (r, D) = 1- \frac {3r} {2D}  + \frac {1} {2} \left( \frac {r} {D} \right) ^3
\label{eq:envelope}
\end{equation}
where $D$ is the diameter of the nanocluster, and $f(r,D) = 0$ for $r>D$. This allows a  simulated PDF to be compared with the results of the X-ray total scattering experimental data.




\subsection{PDFgui modelling }
We used PDFgui \cite{Farrow2007} to fit theoretical three-dimensional structures to the atomic PDF data.
For nanocrystals this approach is based on representing the atomic structure on that of a bulk crystal structure, with thermal motion, and multiplied by an envelope function that simulates the finite length scale. Here the envelope function is that for a spherical nanoparticle as given by equation \ref{eq:envelope}. In this case the diameter $D$ can be treated as a refineable variable, which gives a route to determine the nanoparticle size. Within the refinement the other parameters that are fitted include the lattice parameters, atomic positions and anisotropic atomic displacement parameters, parameters that describe correlated atomic motion, and parameters that model experimental factors that may affect the data. 
 

\section{PDF studies of regular quantum dots of cadmium sulphide}

\subsection{Simulated PDFs} \label{sec:simulated_pdfs}

\begin{figure}[t] 
\begin{center}
\includegraphics[width=8cm]{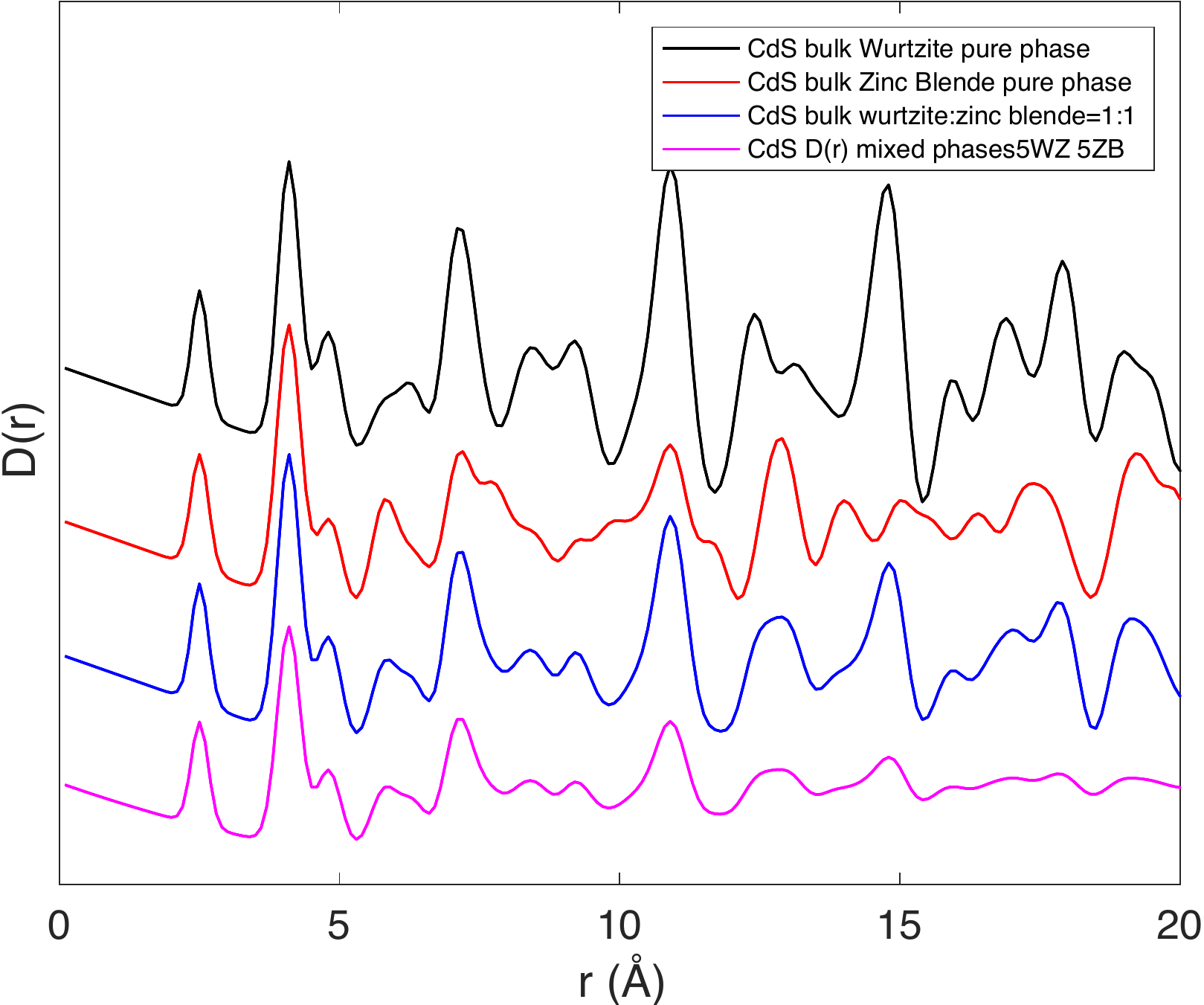}
\caption[]{The  simulated PDFs of two different CdS bulk structures, wurtizite (top curve) and zinc blend (second curve down), compared with a 50\% mixture of the two (third curve down) and of the 50\% mixture multiplied by an envelope function (equation \ref{eq:envelope}) corresponding to a spherical nanocluster of diameter 20 \AA ~(bottom curve).}
\label{fig:simulated_gulp} 
\end{center}
\end{figure}

The value of simulating the PDF is that it gives a noise-free impression of the PDF in advance, enabling us to identify key features. In the present case, it allows us to identify the differences between CdS crystals in either the zinc blend (ZB, cubic space group $F\overline{4}3m$) or wurtzite (WT, hexagonal  space group $P6_3mc$) structures. The simulations of these two phases are compared in Figure \ref{fig:simulated_gulp}. Although the first three peaks are identical, the PDFs are different after around 5~\AA, with the WT PDF having some more pronounced features. For example, the WT PDF has a strong double peak at around 8~\AA, the distance at which there is a clear dip in the PDF of the ZB phase, and a strong peak at around 11~\AA\ where there is a much weaker peak in the PDF of the ZB phase.





In Figure \ref{fig:simulated_gulp} we also show a PDF obtained as a 50:50 mixture of the PDFs of the two crystalline phases to resemble a disordered crystal, and finally we multiply this PDF by an envelop function (equation \ref{eq:envelope}) with $D=20$~\AA\ to simulate a disordered nanoparticle. In the nanoparticle the key signatures of the presence of the WT phase are the double peak at 8~\AA\ and the strong feature at 11~\AA. It can be seen that beyond this point the features in the PDF become weak and we are unable to discriminate the presence or otherwise of features of either phase.

The mixture of the PDFs of the two phases is a reasonably good description of a system with large disorder in the stacking sequence; a previous PDF study of RQDs of CdSe reported the existence of stacking disorder \cite{Masadeh2007}. Both structures contain layers of tetrahedra in a hexagon arrangement. When stacked the layers are displaced laterally to give the bond connections between layers. There are three choices of lateral displacement which we denote as A, B and C. In the WT structure, the layers are stacked in sequence ABAB... along the hexagonal $c$ axis, whereas in ZB the stacking sequence is ABCABC... along the $[111]$ direction. If the stacking sequences in nanoparticles are disordered, we may get quite random sequences, providing nearest neighbours are different. If the start layer is A, the second must be B (C is equivalent, as it will be in each case when we relate to A), then the third layer can be A or C and this is picked up by the mixture of the WT and ZB phases respectively. Similarly the 4th layer may be A or B, and these are also picked up from the ZB and WT phases respectively. It is only when we come to the fifth layer that the mixture always has the A layer, but by the time we get to the fifth layer (distance 28~\AA) we are now outside or near the diameter of the nanoparticle and thus this is not important. What this demonstrates is that by simulating either within a GULP model or in PDFfit using a mixture of two phases we can simulate random stacking sequences in the NCs to a reasonable extent.

\subsection{Experimental data for regular quantum dots and the bulk phase}

The total scattering functions measured for bulk CdS, RQD-439 and RQD-355 are shown in Figure S3. The CdS bulk has sharp peaks (although the diffractometer has a relatively low resolution because it has a short wavelength and is optimised for intensity), which are consistent with the hexagonal wurtzite crystal structure. On the other hand, the total scattering from the two CdS RQDs have broader peaks, and the fine detail fades on increasing $Q$.

\begin{figure}[t] 
\begin{center}
\includegraphics[width=9cm]{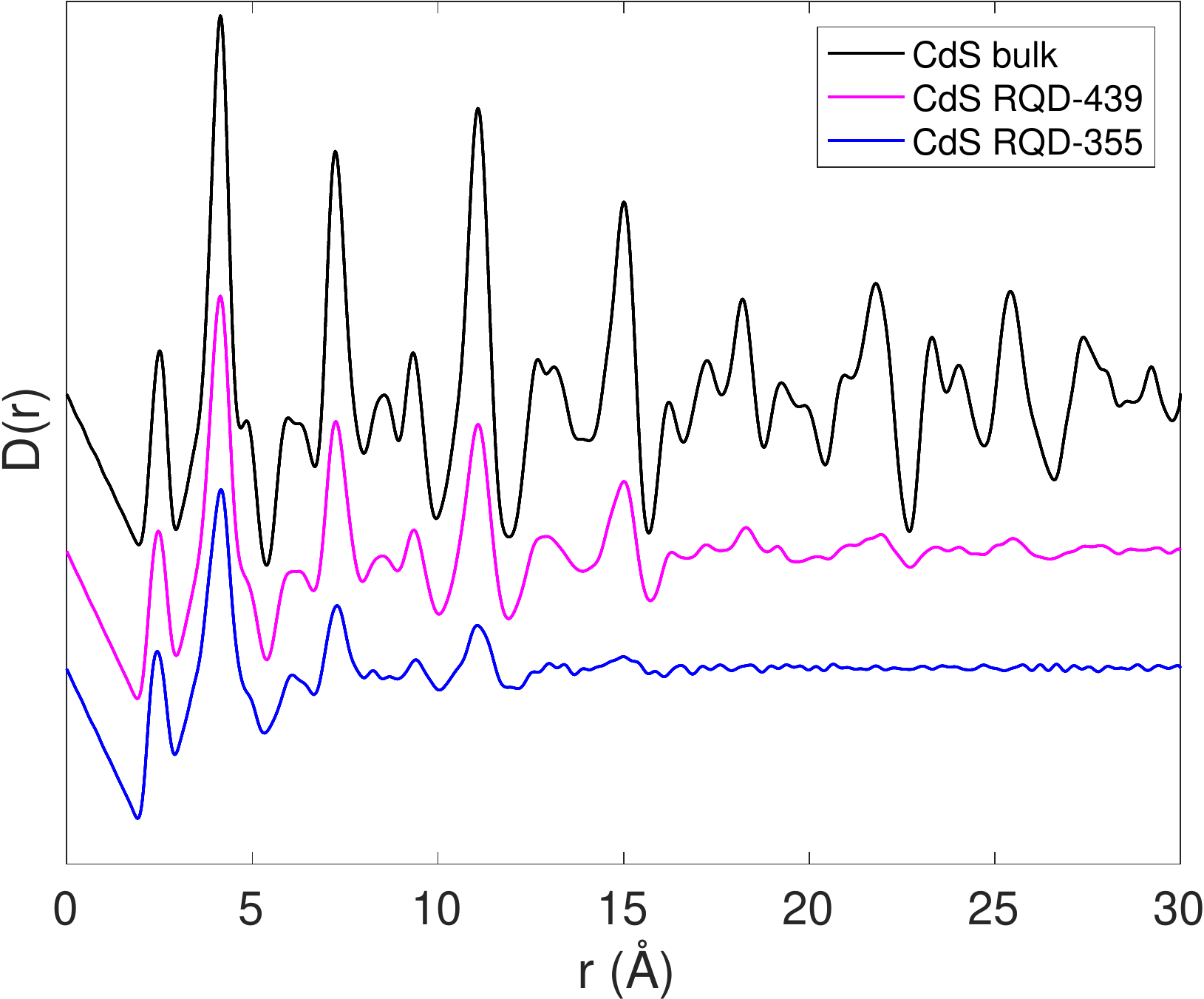}
\caption[]{Comparison of the PDF $D(r)$ of CdS RQD-355 (bottom) and RQD-439 (middle) with that of the bulk phase (top). The key features of the bulk phase up to a distance of around 15 \AA\ are also found in the PDFs of the RQDs. }
\label{fig:cds_rsnc_dr} 
\end{center}
\end{figure}

The PDFs of the RQDs are shown in Figure \ref{fig:cds_rsnc_dr}. The three functions show many similarities to each other, and in turn are very similar to the simulated PDFs shown in Figure \ref{fig:simulated_gulp}. As discussed in Section \ref{sec:simulated_pdfs}, there are characteristics within the PDF that can be identified with the WT and ZB phases, and in the case of the PDF of the bulk it can be seen by inspection to be dominated by features of the WT phase, consistent with the crystal structure analysis. The striking thing is the similarities between the PDFs of the RQDs and the bulk, not only in the first three peaks that define the local structure, but also across the full range of data up to the distance reflecting the effective diameter of the NCs. This implies that the atomic structures of the RQDs are very similar to that of the bulk, based on tetrahedral atomic coordination with stacking of layers as discussed in Section \ref{sec:simulated_pdfs}. We also see clearly that the diameter of RQD-439 is larger than that of RQD-355, which is consistent with the longer wavelength of the UV-vis absorption peak \cite{Zhu2017,Yu2009}, as shown in Figure S1.

Although methods used to generate the PDF data in Figure \ref{fig:cds_rsnc_dr} (as well as in later figures) have removed noise from the region 0--2 \AA, we performed analysis where we retained the noise in this region. Specifically we checked carefully for peaks in the PDF at distances between 1.1--1.6 \AA, which would correspond to C--C and C--O distances in the ligands. The noise in the data were below around 1 \AA, and from this analysis we conclude that we are not seeing pair correlations from within the organic ligands in our experimental PDFs. This of course is not surprising given that peaks in the PDF are weighted by the products of atomic numbers, which are much larger for cadmium and sulphur. This is consistent with data published in reference \onlinecite{Zhang:inpress}, and with a previous PDF study of CdSe \cite{Masadeh2007}.


The three PDFs of Figure \ref{fig:cds_rsnc_dr} were analysed using PDFgui\cite{Farrow2007}, fitting the experimental PDFs with calculated functions based on a mixture of both WT and ZB structures, treating the phase fraction as a variable in the fitting process, together with the parameters of the crystal structures. 
In fact we tried fitting to single phases as shown in Figure S4, but in each case the quality of the fit was worse, indicated by comparing the fitted and experimental PDFs by eye, and quantified by the goodness-of-fit parameter $R_\mathrm{w}$ as defined in reference \onlinecite{Masadeh2007}.

\begin{figure}[t] 
\begin{center}
\includegraphics[width=9cm]{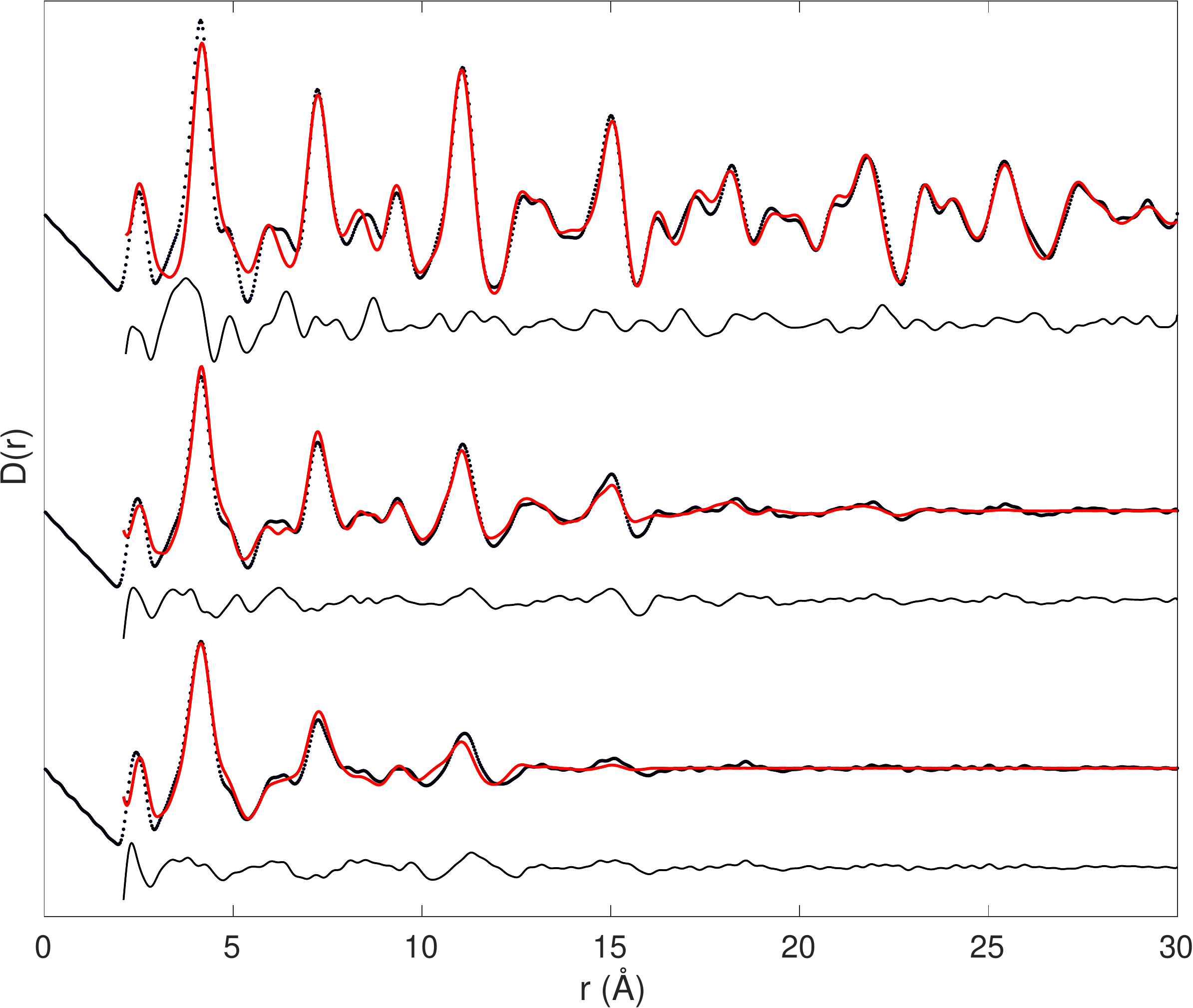}
\caption[]{Fitted PDF data $D(r)$ for the bulk phase (top curve) and nanocluster phases RQD-439 (middle) and RQD-355 (bottom). The experimental data are the circles and the fitted functions are the continuous curves. All data are fitted with a mixed phase containing the wurtzite (space group $P6_3mc$) and the zinc-blende (space grosup $F\overline{4}3m$) structures.}
\label{fig:pdfgui_mixedphase} 
\end{center}
\end{figure}

\begin{table}
\caption{The refined parameter values for the atomic structures of the three forms of CdS as obtained from PDF analysis by PDFgui, modelling structural disorder by using a mixed phase of the wurtzite (WT, hexagonal space group $P6_3mc$) and zinc blende (ZB, cubic space group $F\overline{4}3m$) structures.} 
\label{table:mixed}
\begin{tabular}{@{}llllll} 
\hline \hline
 & Bulk & RQD-355& RQD-439 \\
\hline
\% WT & 69.92(8) & 35(14)  & 66(28) \\
ZB $a$ (\AA) & 5.9164(9) & 5.92(9) &5.92(8) \\
WT $a$ (\AA)& 4.1889(6)&4.1(3)&4.18(6)\\
WT $c$ (\AA) & 6.8217(9)&6.8(5) &6.9(1) \\
$D$ (\AA) & ---  & 17(4)  & 28(4) \\ 
$R_\mathrm{w}$ & 0.24 & 0.26 & 0.24\\
\hline \hline
\end{tabular} 
\end{table}


The experimental and fitted PDFs are compared in Figure \ref{fig:pdfgui_mixedphase}, and values of the refined parameters are given in Table \ref{table:mixed}. The two RQDs have relatively small diameters as obtained in the fitting, $28\pm4$~\AA\ for RQD-439 and $17\pm4$~\AA\ for RQD-355. The refined values of the phase fractions are different in each case; however, it should be noted that these values have relatively large errors for the two nanoclusters, whereas there is a much smaller error for the bulk. We found that the refinement of the diameter of nanoparticle is highly correlated with other parameters such as anisotropic ADPs, phase factors, and correlated motion parameters \cite{Farrow2007}. Thus the refinement result is not as robust as usually found for crystalline materials.
In this case we can identify some possible reasons for this. First is that the differences between the PDFs of the two crystalline phases is not large at small $r$, as we discussed in Section \ref{sec:simulated_pdfs}, and may be smaller than the difference between the experimental and simulated PDFs at low $r$. Second, the description of the PDF as that of a bulk with a shape function neglects effects at the edge, which is more important for smaller NCs. In particular, we are likely to have more Cd atoms at the edge relative to the bulk because of the coating with negatively-charged ligands. In turn this may lead to inconsistencies within the fitting, for example with the number of Cd--S neighbours being different than assumed by this representation of the NC structure. 
We also found that the values of the lattice parameters of two phases when fitted together can have large errors. To a large extent this can be understood through correlations with the peak width parameters. The lattice parameters are largely determined by the Cd--S distance and thus will lead to the same first two peaks in the PDF. A small divergence of the lattice parameters will broaden these peaks, which is the correlation we have noted. There are too few additional peaks to act as a constraint.
The value of the NC diameter $D$ is determined by the high-$r$ data.  However, at high $r$ the PDF is relatively small. In the refinement of the model structure, the  least squares algorithm may pay more attention to the low-$r$ area because that is where the discrepancies between model and data are largest. These discrepancies are larger than the structure in the PDF at large $r$, which means that the least-squares refinement can effectively ``choose'' to ignore the large-$r$ data. As a result, we end up with a relatively large error on the diameter.

\section{PDF studies of magic-size clusters}

\subsection{CdS MSC-311}

\begin{figure}[t] 
\begin{center}
\includegraphics[width=9cm]{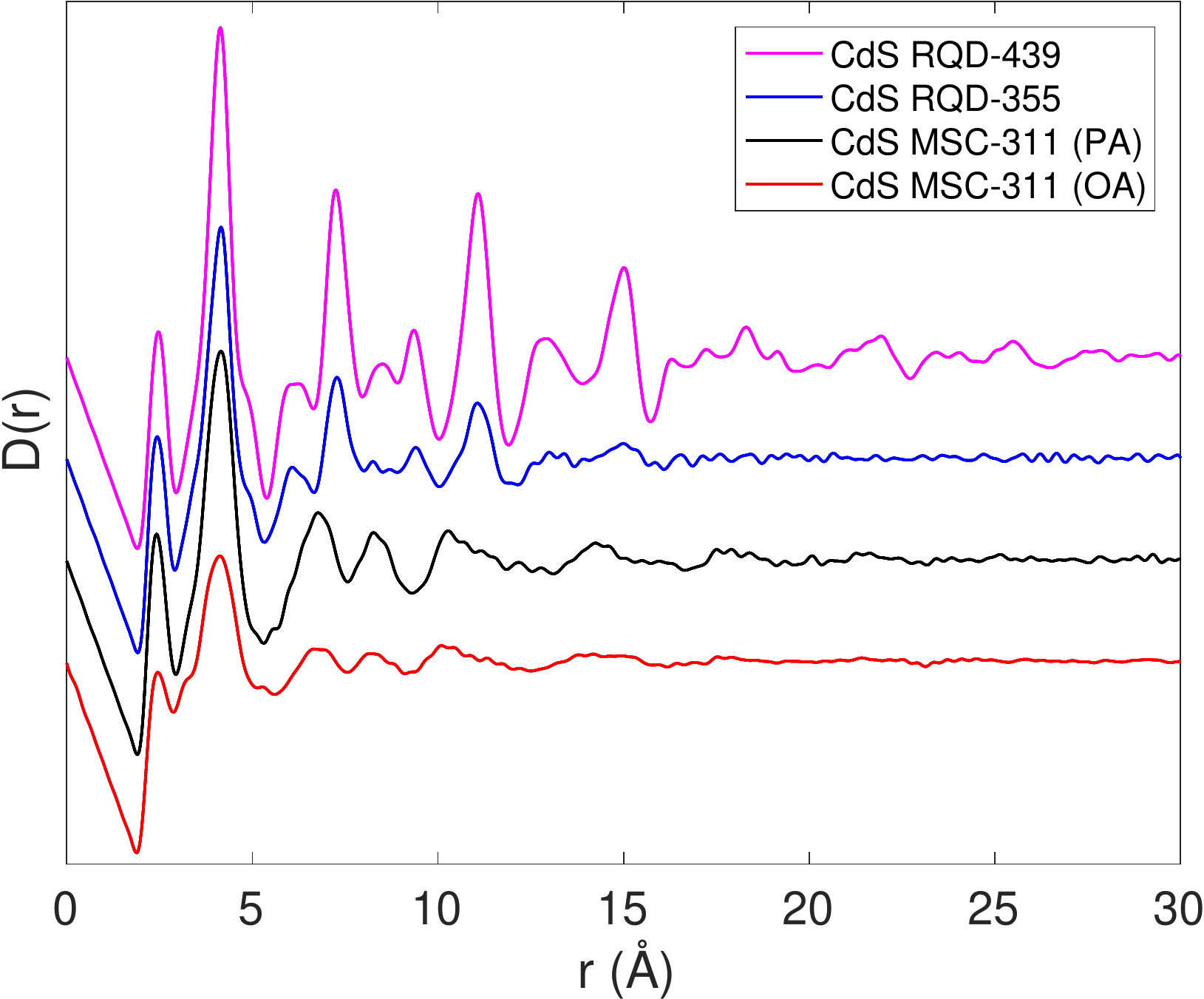}
\caption[]{The PDFs $D(r)$ of the MSC-311 samples (OA MSC-311, bottom, and PA MSC-311, second bottom) compared with those for the RQDs (RQD-439, top, and RQD-355, second top) nanoclusters. The important point is that the PDFs of the RQDs and MSCs have the same features up to a distance of about 5 \AA, but beyond that the PDFs show different features.}
\label{fig:cds311} 
\end{center}
\end{figure}

The PDFs of both OA MSC-311 and PA MSC-311 are shown in Figure \ref{fig:cds311}, where they are compared to the PDFs of RQD-355 and RQD-439. The PDFs of both MSC-311 samples are very similar in terms of the positions, widths and relative heights across the full range of distances within the PDF, and also with respect to the decrease in the PDF amplitude on increasing distance. Our data in Figure \ref{fig:cds311} are also consistent with separate measurements for PA MSC-311 on different samples as reported in reference \onlinecite{Zhang:inpress}, and, importantly, we have reproduced the PDFs from measurements on several different samples ourselves.

Up to 5~\AA\ the PDFs of the MSCs and RQDs look very similar. This means that the local coordination of the atoms in the MSCs are very similar to those in the RQDs, both in terms of interatomic distances and numbers of neighbours. This effectively precludes the possibility that the atomic structure of these MSCs has the form of a shell reminiscent of C$_{60}$ \cite{HAWKINS1991}, because the coordination number of atoms in a shell is 3 rather than 4 in the bulk, which the local atomic structures of the RQDs resemble.

Beyond 5 \AA, the PDFs of MSC-311 and the RQDs look very different. For example, in the range of 5--9 \AA, the CdS RQDs have four peaks in their PDFs, while the PDFs of the MSCs display two peaks only, and without any correspondence of peak position. It is clear therefore that CdS MSC-311 has a significantly different atomic structure from that of the RQDs beyond the first 2--3 neighbours. 

\begin{figure}[t] 
\begin{center}
\includegraphics[width=9cm]{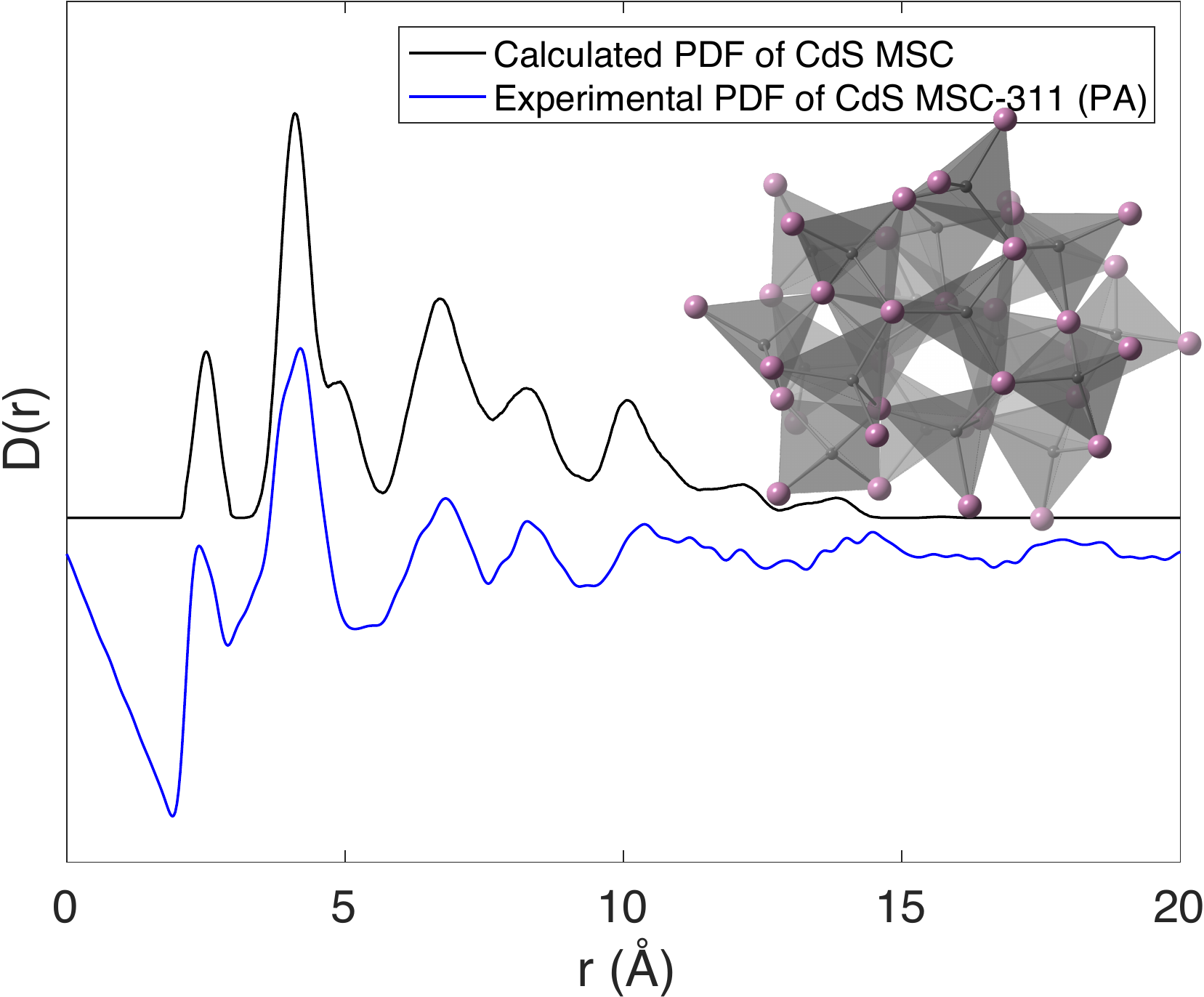}
\caption[]{Comparison of the PDF function $D(r)$ of MSC-311 with a calculated distribution function (essentially a histogram of interatomic distances scaled by the product of the atomic numbers and broadened by a Gaussian function) from the published atomic structure of InP MSC structure, replacing In by Cd and P by S. The atomic structure is shown as the inset (from reference \onlinecite{Gary2016}).}
\label{fig:InP} 
\end{center}
\end{figure}

It is instructive to compare our results with the atomic structure of the  single-crystal assemblage of InP MSCs \cite{Gary2016}. This consists of corner-sharing PIn$_4$ tetrahedra, and within the core each In atom has four P neighbours as in the bulk phases. The cluster has formula P$_{20}$In$_{37}$, with the excess of In cations being on the outside of the cluster to ensure that every P atom has tetrahedral environment. The In cations on the edge of the NC are available to bond to the negatively-charged ligands. Unlike in the bulk and RQDs, the PIn$_4$ tetrahedra are not arranged in layers. Based on replacing the In and P atoms by Cd and S respectively (we did not need to change the interatomic distance), and neglecting the ligands since they are negligible in the x-ray total scattering data, we have calculated the histogram of interatomic distances. This is compared with the $D(r)$ function for PA 311 in Figure  \ref{fig:InP}, where the structure is shown as an inset to the figure. It is striking that the key features that differentiate the PDFs of MSC-311 and of the RQDs, namely the peaks after 5 \AA, are reproduced by the InP structure. Thus we propose that the InP structure, with corner-sharing SCd$_4$ tetrahedra arranged in a manner that is different from that of the bulk phases, is a good initial model for the structure of MSC-311. The mass of this cluster is 4800 Da, excluding any contribution from the ligands. The mass of CdS MSCs has been measured as 5160 Da.\cite{Zhang:inpress} The small difference (7.5 \%) may be accounted for by addition of one S and three Cd atoms to the InP-based nanocluster.

We can see from Figure \ref{fig:cds311} that the decrease in the amplitudes of the features in the PDFs with increasing $r$ is between that of the two RQDs. This means that the effective diameter of MSC-311 is between those of RQD-355 (17 \AA) and RQD-439 (28 \AA). From comparison to the results for the RQDs given in Table \ref{table:mixed}, and noting that visually it appears that the diameter is closer to that of RQD-355 than RQD-439, we estimate a value of round $20 \pm 4$ \AA\ for the diameter of MSC-311. This is larger than the largest Cd--Cd distance in a NC based on the InP structure (15.7 \AA), which is consistent with our suggestion above that the MSC-311 may have slightly more atoms in the InP MSC.

It is also important before we consider MSC-322 to re-iterate that the PDF measurements have shown that the atomic structures of OA MSC-311 and PA MSC-311 have effectively identical atomic structure, in spite of the fact that they are coated with different ligands.

\subsection{CdS MSC-322}

\begin{figure}[t] 
\begin{center}
\includegraphics[width=9cm]{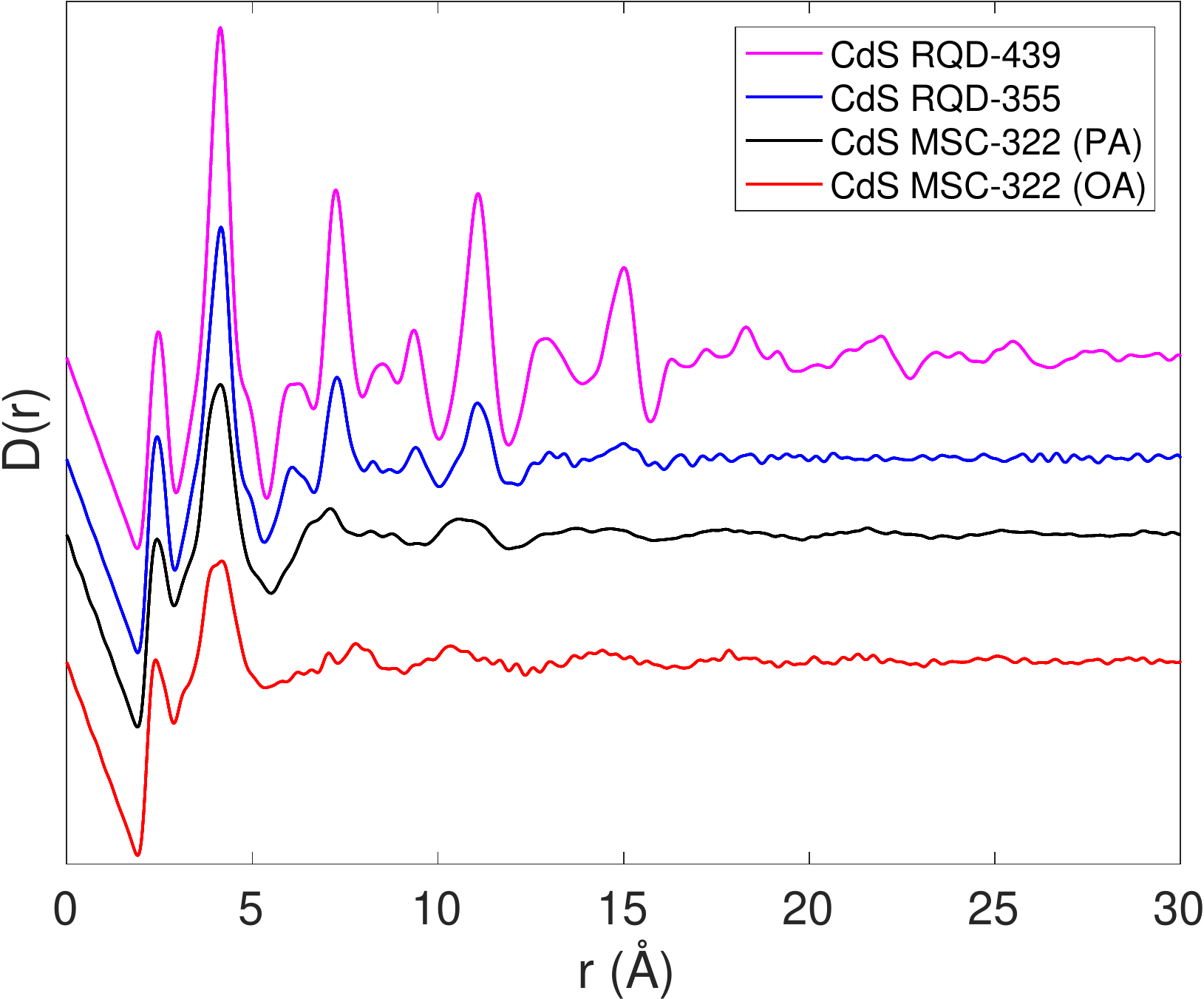}
\caption[]{The PDFs $D(r)$ of the magic-size (OA 322, bottom, and PA 322, second bottom) and regular (RSN 439, top, and RSN 355, second top) nanoclusters. }
\label{fig:cds322} 
\end{center}
\end{figure}

The PDFs of OA MSC-322 and PA MSC-322 are shown in Figure \ref{fig:cds322}, and again they are compared with the PDFs of RQD-355 and RQD-439. Our data for PA MSC-322  presented here are consistent with the data for PA MSC-322 obtained by independent measurements on a different sample in reference \onlinecite{Zhang:inpress}. We have reproduced the data for OA MSC-322 given in Figure \ref{fig:cds322} using three different samples.

As in the case of MSC-311, the PDFs of the MSC-322 samples are similar to those of the RQDs up to a distance of 5 Å, and by comparison with Figure \ref{fig:cds311}  they also similar to the PDF of MSC-311 over this range of distances.
Similar to MSC-311, the PDFs for both samples of MSC-322 are significantly different from the RQD PDFs for distances above 5 \AA. What is particularly interesting in Figure \ref{fig:cds322} is the differences between the PDFs of OA MSC-322 and PA MSC-322 for distances between 5--9 \AA. In particular, PA MSC-322 has a peak at 7 \AA\ and OA MSC-322 has a peak at 8 \AA. The case of PA MSC-322 is interesting because its PDF has many similarities with MSC-311, with the only significant difference being with regard to the peak at 8 \AA\ in MSC-311.

\section{Discussion and conclusions}

In this paper we have presented an X-ray total scattering study of various nanoclusters of CdS, focussing analysis on the PDFs derived from the scattering data. The results are sensitive primarily to the Cd and S ions, with the organic ligands showing no discernible features in the PDF. We have compared the PDFs of the nanoclusters with that for the bulk, and compared the PDFs of regular quantum dots with magic-size clusters. The results present an interesting contrast between regular and magic-size clusters. 

Measurements on two CdS RQDs of different diameter have shown that their atomic structures are very similar to the atomic structures of the bulk phases, with disorder in the stacking of the layers found in the crystalline phases. We have obtained estimates of their effective diameter through fitting an atomic model to the PDF data assuming that the atomic structure is based on a disordered stacking sequence of the tetrahedral layers found in the two bulk phases, wurtzite and zinc blend, and that the effect of the finite sizes of the NCs can be represented by a multiplicative spherical shape function. We obtained diameter values of $17 \pm 4$ and $28 \pm 4$ \AA.


Whilst the CdS RQDs have atomic structures that resemble the bulk phases of CdS, it is clear that the atomic structures of the MSCs are only similar over short distances. This implies that the atomic structures of the MSCs are  based on the same tetrahedral arrangements of atoms, but that their longer-range structure departs from the layer structures of the bulk phases that are also found in the RQDs. This result excludes the possibility that the MSCs have qualitatively different types of atomic structures such as cage structures  with equal numbers of cations and anions\cite{Kasuya2004, Tohgha2013, Nguyen2010}. A recently published structure of the MSC of InP shows that it is possible to form structures consistent with this conclusion, and indeed the PDF from a model for CdS based on the InP structure shows remarkable agreement with the PDF for the MSC-311, particularly with regard to the features that are different in the PDFs of the RQDs. Thus we conclude that the MSCs have atomic structures that consist of arrangements of corner-sharing SCd$_4$ tetrahedra that are dissimilar to the stacking within layers as seen in the bulk and RQDs, with Cd cations on the surface of the MSCs.

One interesting result is that MSC-311 has the same atomic structure with both PA and OA ligands as coatings, but that, by contrast, the atomic structure of MSC-322 appears to be different for the two ligands. The short-range structure of both types of MSC-322 is again similar to that of the RQDs and MSC-311, suggesting the same structural base of corner-sharing SCd$_4$ tetrahedra, but the exact arrangement is different in the two cases albeit with a very similar band gap.

The challenge now is to use the PDF data to develop detailed models for the atomic structure of these MSCs. In the absence of single crystals we will need to use advance modelling techniques such as the Reverse Monte Carlo method; we are now tackling this challenging problem.

\begin{acknowledgments}
LT and LF are grateful for financial support from the Chinese Scholarship Council and Queen Mary University of London. 
\end{acknowledgments}

\bibliography{firstpaper_lei_mtd2_1.bib}

\end{document}


\frenchspacing
\title[Chalcogenide nano clusters]{X-ray total scattering study of regular and magic-size nanoclusters of cadmium sulphide: supporting information}
\author{Lei Tan,  Alston J Misquitta, Andrei Sapelkin}  
\affiliation{School of Physics and Astronomy, Queen Mary University of London, Mile End Road, London, E1 4NS,UK}
\author{Le Fang}
\affiliation{School of Biological and Chemical Sciences, Queen Mary University of London, Mile End Road, London, E1 4NS, UK}
\author{Rory M Wilson}
\affiliation{School of Engineering and Materials Science, Queen Mary University of London, Mile End Road, London, E1 4NS, UK}
\author{Baowei Zhang, Tingting Zhu, Frank S Riehle, Shuo Han, Kui Yu}
\affiliation{Institute of Atomic and Molecular Physics, Sichuan University, Chengdu, 610065, P. R. China}
\author{Martin T Dove}
\email[Corresponding author: ]{martin.dove@qmul.ac.uk}
\affiliation{School of Physics and Astronomy, Queen Mary University of London, Mile End Road, London, E1 4NS, UK, and School of Physics, Sichuan University, Chengdu, 610065, P. R. China}

\maketitle

\section{sample synthesis and characterization}
\subsection{Sample synthesis}

To prepare the CdS MSC-311 and MSC-322 with phenylacetic acid (PA) ligands, the Cd precursor Cd(OOCCH$_2$Ph)$_2$ -- Ph being the phenyl group C$_6$H$_5$ -- was prepared by mixing Cd(OCOOH)$_2$  (0.60 mmol) and PhCH$_2$COOH (2.16 mmol) in a temperature at 90~$^{\circ}$C. The reaction mixture then heated to 90~$^{\circ}$C, to which 1 mL of toluene was added. The Cd precursor was maintained at 60~$^{\circ}$C for the preparation of MSC-311, or 90~$^{\circ}$C for the preparation of MSC-322\cite{Zhang:inpress}. 
A mixture of Bis(trimethylsilyl) sulphide (0.30 mmol), and toluene (1 mL) was then injected into the solution under a nitrogen atmosphere which was held at the injection temperature for about 5 minutes. 
After the reaction had finished, the mixture was then cooled rapidly in an ice-water bath, resulting in the production of MSCs.
For the purification of the CdS MSCs, the as-synthesised reaction solution was centrifuged directly. The precipitate was then washed twice by ethyl acetate (5ml) to remove the unreacted precursors. The precipitate was quickly dried under vacuum.

The synthesis method of CdS MSC-311 and 322 with oleic acid (OA) ligands followed the method of Nevers et al\cite{Nevers2017,Nevers2018}. 
Breifly, the cadmium precursor Cd oleate was prepared by heating a mixture of CdO (10 mmol) and oleic acid (31.6 mmol) to 160 $^{\circ}$C under nitrogen atmosphere. After the mixture formed a tan clear solution, it was then cooled down to 50 $^{\circ}$C for the next step which sulphide precursor was injected.  
The sulphide precursor tri-n-octylphosphine sulphide (TOPS) was prepared by dissolving sulphur (5 mmol) in ri-n-octylphosphine (TOP) (2.0~mL). 
The sulphide precursor was then injected into Cd oleate solution at 50 $^{\circ}$C and heated to 140 $^{\circ}$C. It was then soaked at 140~$^{\circ}$C for 65 mins. 
The reaction was quenched with ethyl acetate, resulting in the formation of MSC-322. For the purification of the CdS MSCs, the as-synthesised reaction solution was centrifuged directly and then washed twice by ethyl acetate (10ml). The precipitate was quickly dried under vacuum.
The MSC-311 was synthesised by the transformation from MSC-322 to MSC-311 \cite{Nevers2017,Zhang:inpress}. 
Briefly,  50 mL of ethanol was added to 100 mg of MSC-322 and then stirred at room temperature (in air) for 24 hrs. Over 24 hrs, the waxy-like MSC-322 sample transformed into a fine white powder as it converted into MSC-311. Interestingly we were unable to see the same direct transformation with the MSC with PA ligands.


\subsection{Sample characterisation}
\subsubsection{UV-vis absorption spectra}
\begin{figure}[t]
\begin{center}
\includegraphics[width=5.5cm]{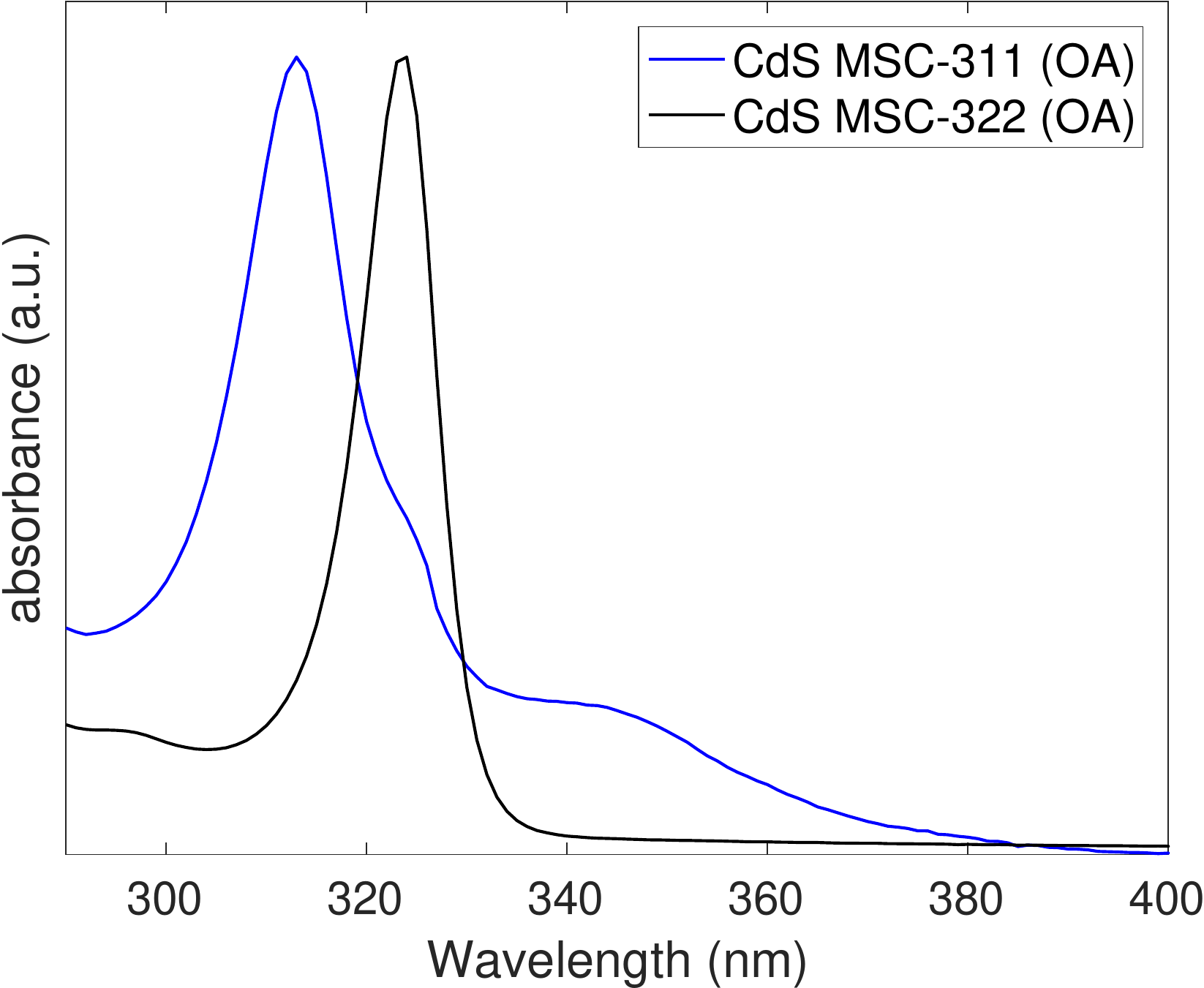}
\includegraphics[width=5.5cm]{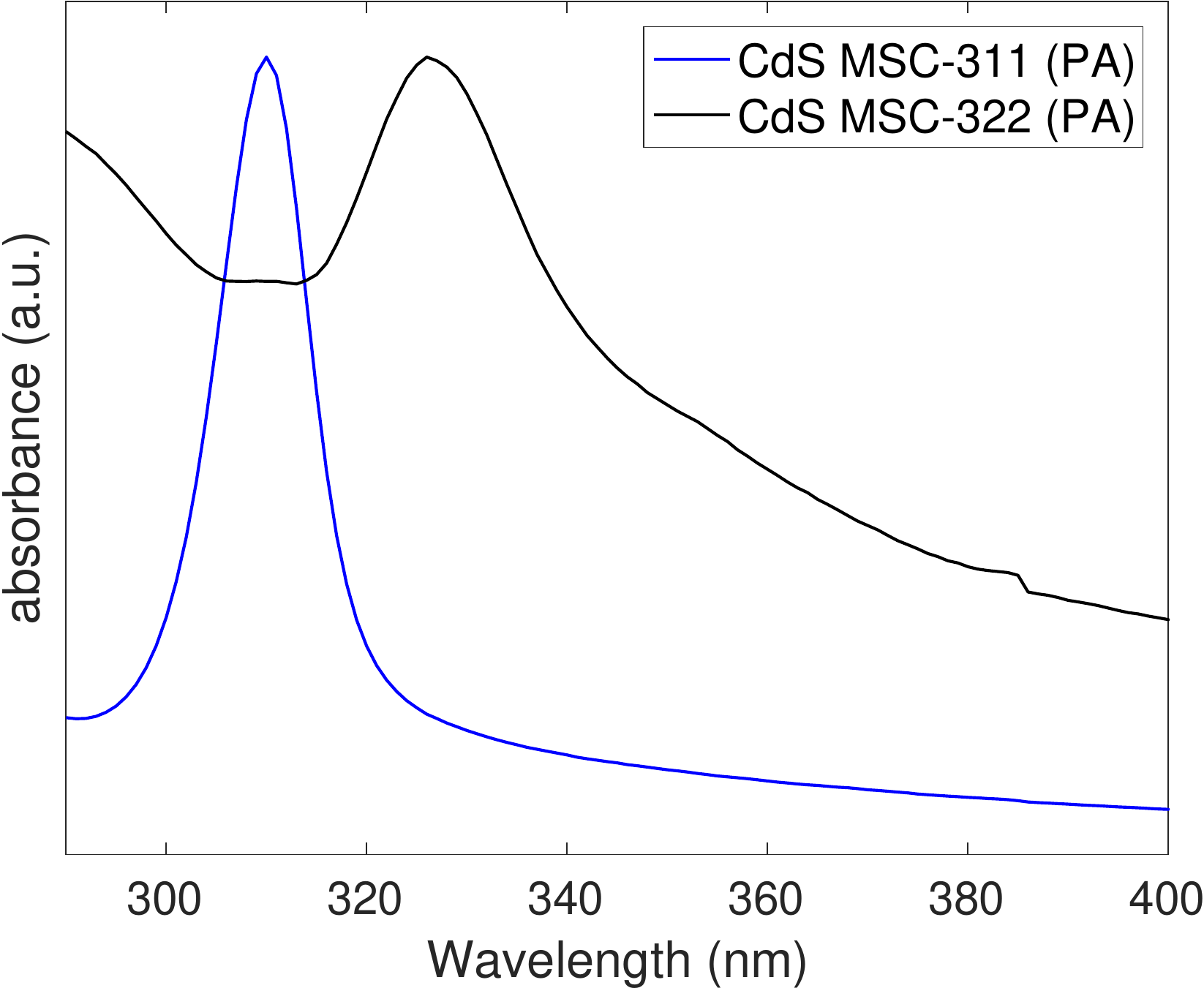}
\includegraphics[width=5.5cm]{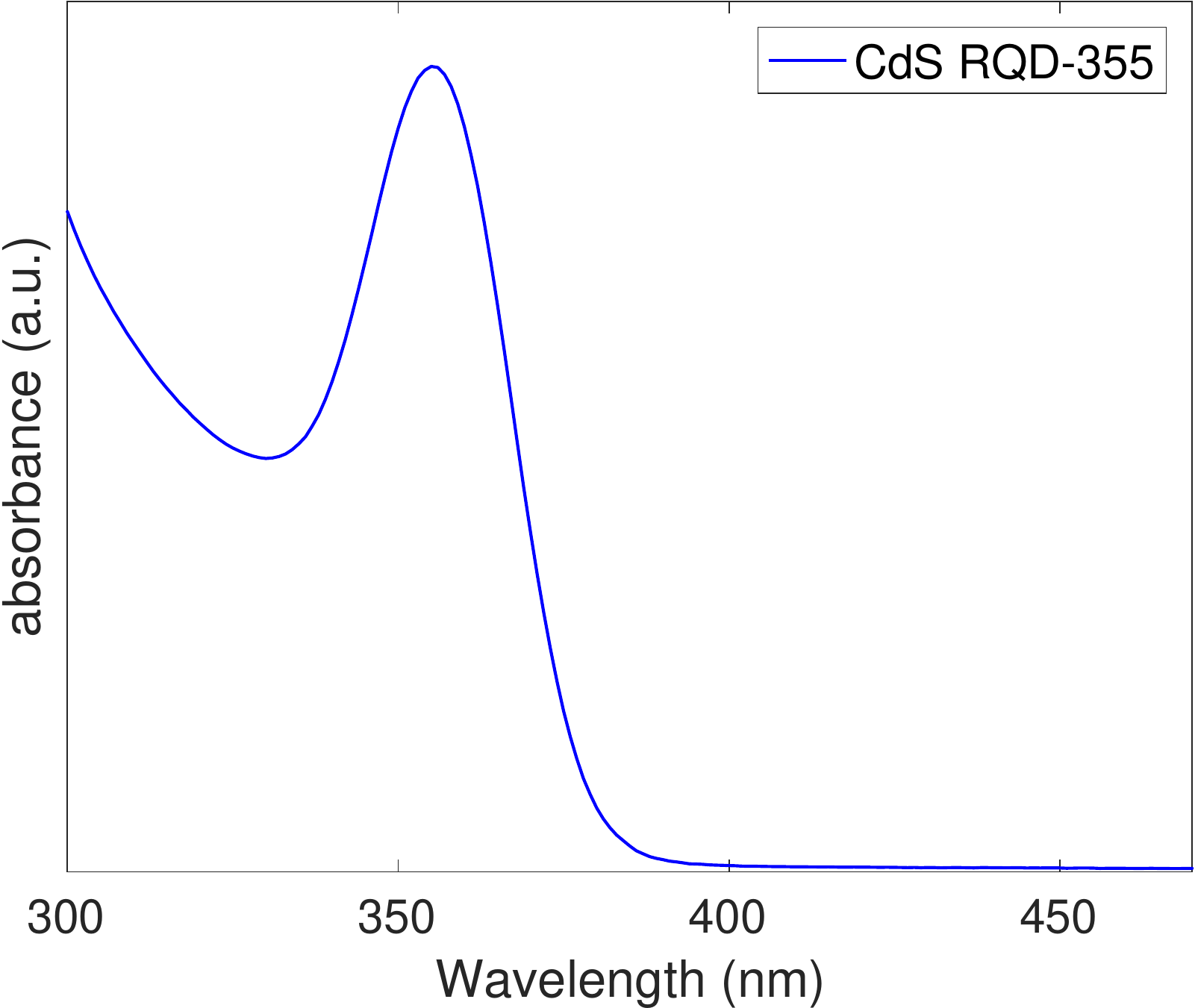}
\includegraphics[width=5.5cm]{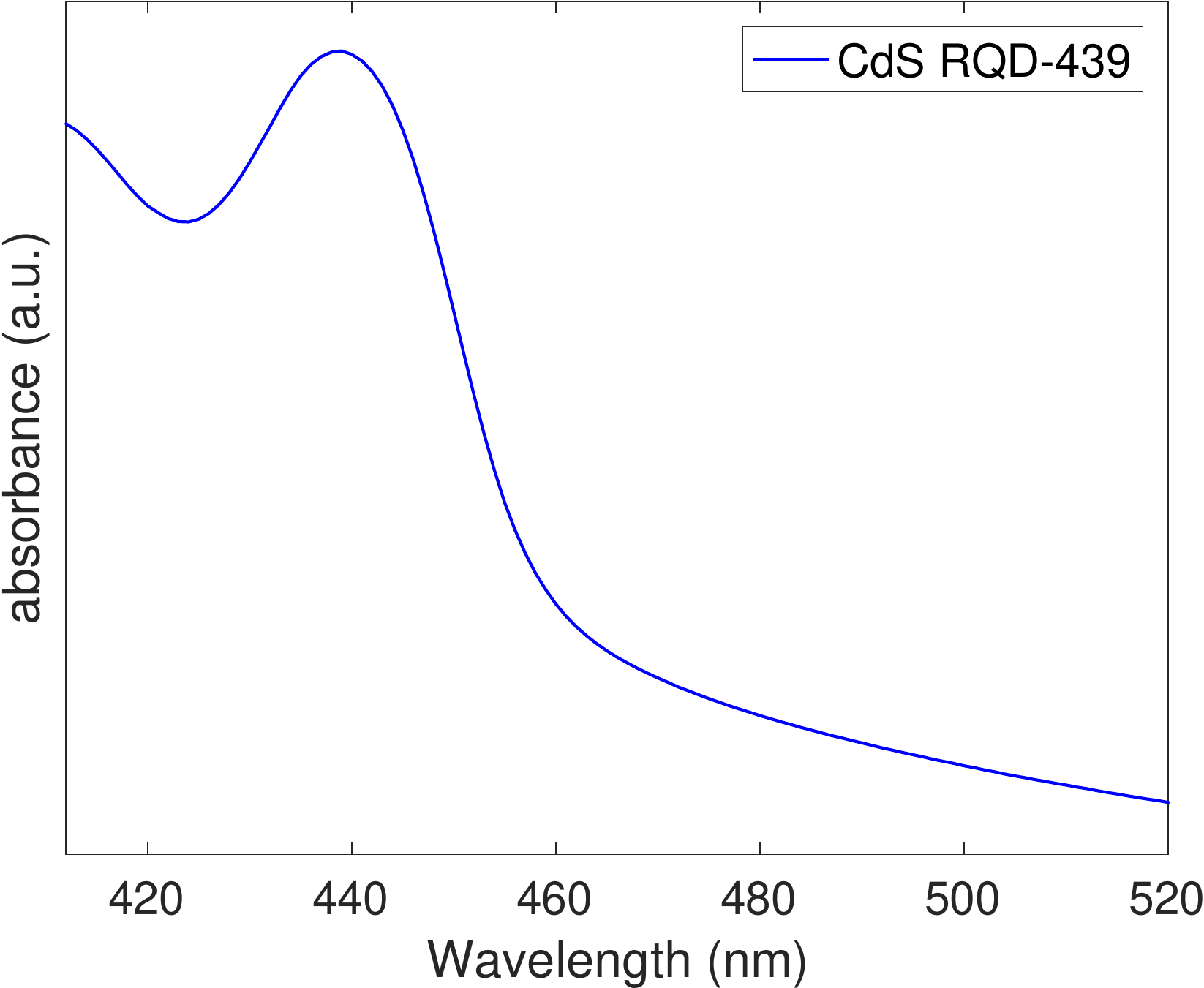}
\caption[]{Optical absorption spectra of CdS MSC-311, MSC-322 of two different ligands (OA ligand and  PA ligand)  and RQD-355, RQD-439, collected from a solution in which the CdS MSCs (ca. 20 mg) dissolved in toluene (3 mL).} 
\label{fig:uvis} 
\end{center}
\end{figure}

The samples were characterised by ultraviolet visible (UV-vis) absorption as shown in Figure \ref{fig:uvis}. The two CdS  MSCs with sharper absorption peaks at 311 nm and 322 nm  when compared to the CdS RQDs, which is attributed to the small and exact size of the CdS MSCs.

\subsubsection{Fourier Transform Infrared Spectroscopy Analysis}
\begin{figure}[t]
\begin{center}
\includegraphics[width=8cm]{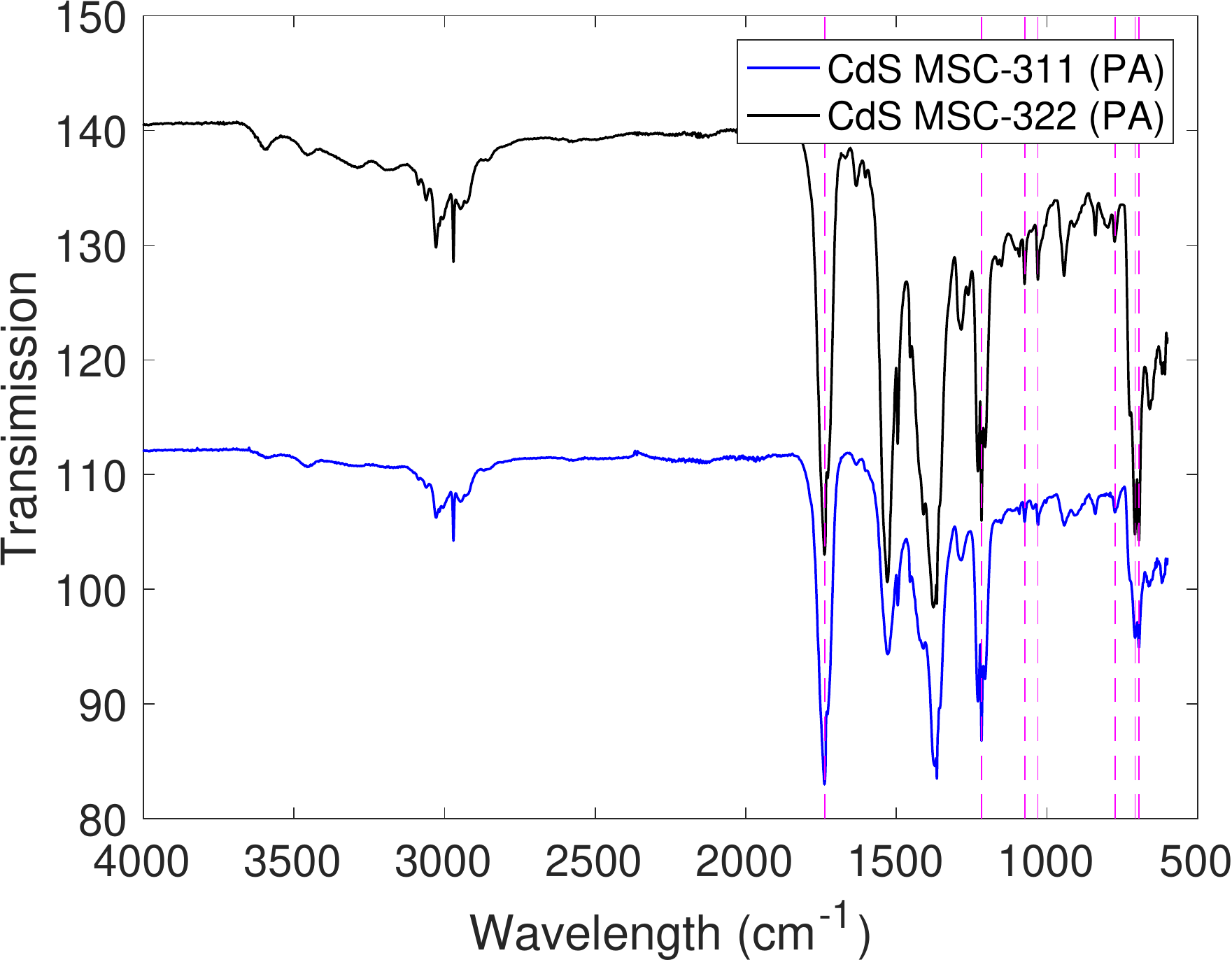}
\includegraphics[width=8cm]{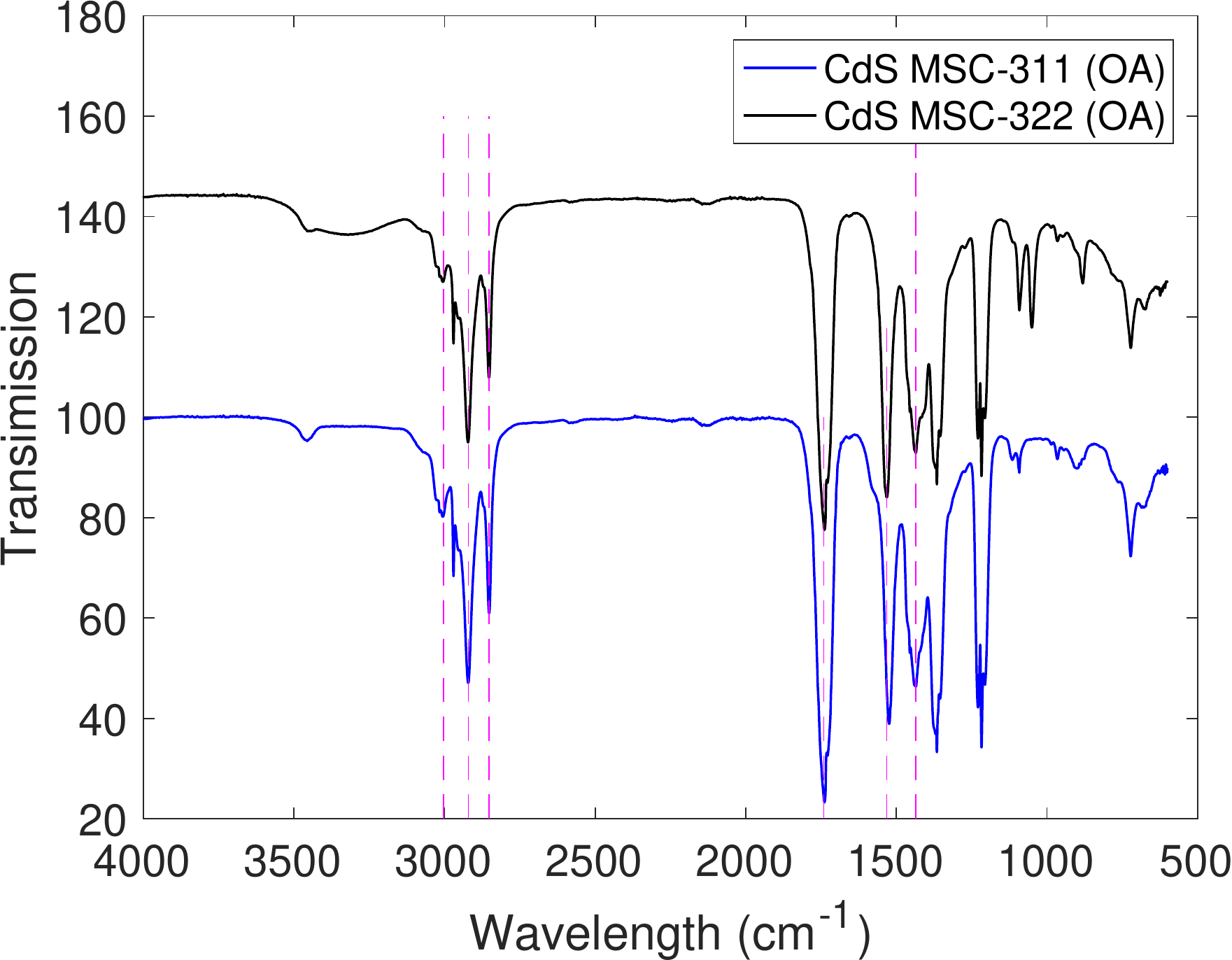}
\caption[]{FTIR spectra for four CdS magic-size cluster coated with two different ligands (OA311, OA322, PA311, PA322). The peak features are highlighted with dash line.}
\label{fig:IR} 
\end{center}
\end{figure}

We have two different kinds of ligands based on the different synthesis precursor we choose as described in the synthesis section above.
The surface ligands of the four magic-size clusters(OA311, OA322, PA311, PA322) are studied by Fourier transform infrared spectroscopy (FTIR) as shown in Figure \ref{fig:IR}. 
The OA311 and OA322 have similar, although not identical spectra, consistent with the spectrum of oleic acid. The small differences may be attributed to bonding on the surfaces of the nanoclusters \cite{Nevers2017,Zhang:inpress}. The 1435 cm$^{-1}$  and 1532 cm$^{-1}$  bands can be assigned to the carboxylate stretches \cite{Nevers2017}.  An absorption band at 1738 cm$^{-1}$ can be related to C=O stretching modes for monomeric COOH species \cite{Lee1999}. Two main peaks at 2925 cm$^{-1}$ and 2852 cm$^{-1}$ are the asymmetrical and symmetrical stretching of CH$_2$, and the shoulder peak at 3003 cm$^{-1}$  is the C--H stretch in the C=C--H alkenyl group \cite{Lin-Vien1991}. All these typical features suggested the formation of oleate ligands attached to the particle surface.

 
Both the PA 311 and 322 show similar FTIR spectra features compare with the phenylacetic acid. 
 As shown in Figure S2, an absorption band was observed in 1738 cm$^{-1}$, which was mainly caused by the stretching of the C=O bond \cite{Badawi2011}. 
 The existence of one or more aromatic rings in a structure is normally readily determined from the C--H and C=C--C ring-related vibrations\cite{Coates2006}. The observed band at 1495 cm$^{-1}$  is associated with the aromatic ring stretch. The three peaks at 1030 cm$^{-1}$, 1074 cm$^{-1}$ and 1217 cm$^{-1}$  can be assigned to the aromatic C--H in-plane bend, and the aromatic C--H out-of-plane bend can be observed with wavelengths of  694 cm$^{-1}$, 707 cm$^{-1}$, 774 cm$^{-1}$ and 840 cm$^{-1}$. \cite{Coates2006}
The FTIR result indicates that the nanoclusters synthesised by Cd(OOCCH$_2$Ph)$_2$ precursor are passivated by a phenylacetic ligand.

\section{Experimental PDF analysis}
\subsection{Experimental PDF compare between CdS  bulk and RQDs }
\begin{figure}[t]
\begin{center}
\includegraphics[width=8cm]{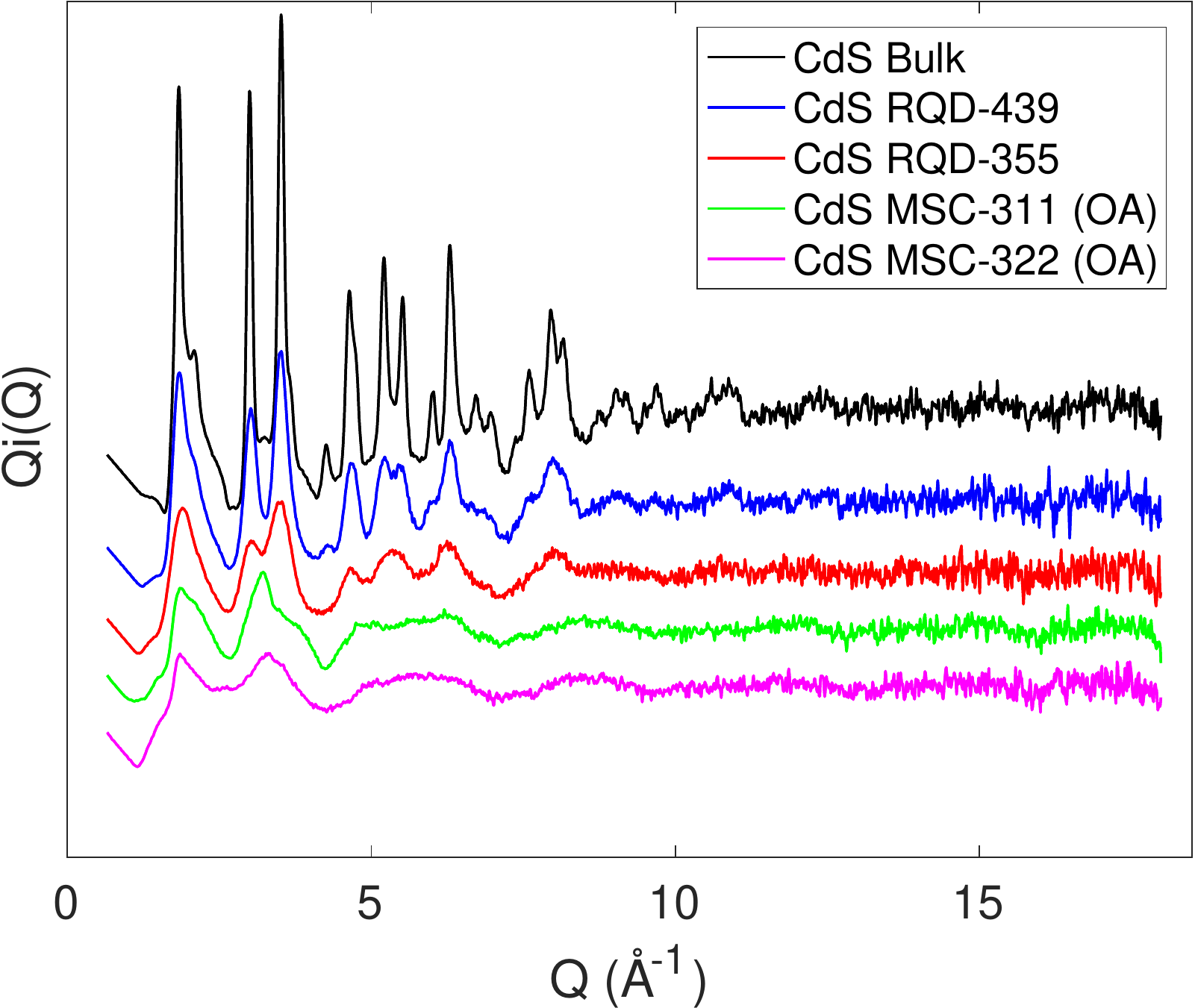}
\caption[]{The total scattering data function $Qi(Q)$ for the bulk phase of CdS (top curve) and the two regular quantum dots RQD-439 and RQD-355 (middle and bottom curves respectively). Note the sharp Bragg peaks in the data for the bulk sample, whose positions are reflected in the broader peaks in the scattering functions of the two RQDs.}
\label{fig:cds_rsnc_QiQ} 
\end{center}
\end{figure}

The total scattering functions measured for bulk CdS and the two CdS regular quantum dots( CdS RQD-439 and CdS RQD-355) are shown in Figure \ref{fig:cds_rsnc_QiQ}. 

\subsection{PDFgui modeling result of CdS  bulk and RQDs }
\begin{table}[t]
\caption{The refined residual $R_\mathrm{w}$ values obtained from PDF analysis assuming the wurtzite and zinc blende structure models with space groups $P6_3mc$ and $F\overline{4}3m$, respectively.} 
\label{table:rw}
\begin{tabular}{@{}llllll}
\hline \hline
 & CdS bulk & CdS RQD-355 & CdS RQD-439\\
\hline
wurtzite ($R_\mathrm{w})  $ &  ~ 0.58 &   ~ 0.36 & ~ 0.36 \\
zinc blend ($R_\mathrm{w}) $ &  ~ 0.33 &    ~ 0.35 &  ~ 0.37\\
\hline \hline
\end{tabular} 
\end{table}

\begin{figure}[t]
\begin{center}
\includegraphics[width=8cm]{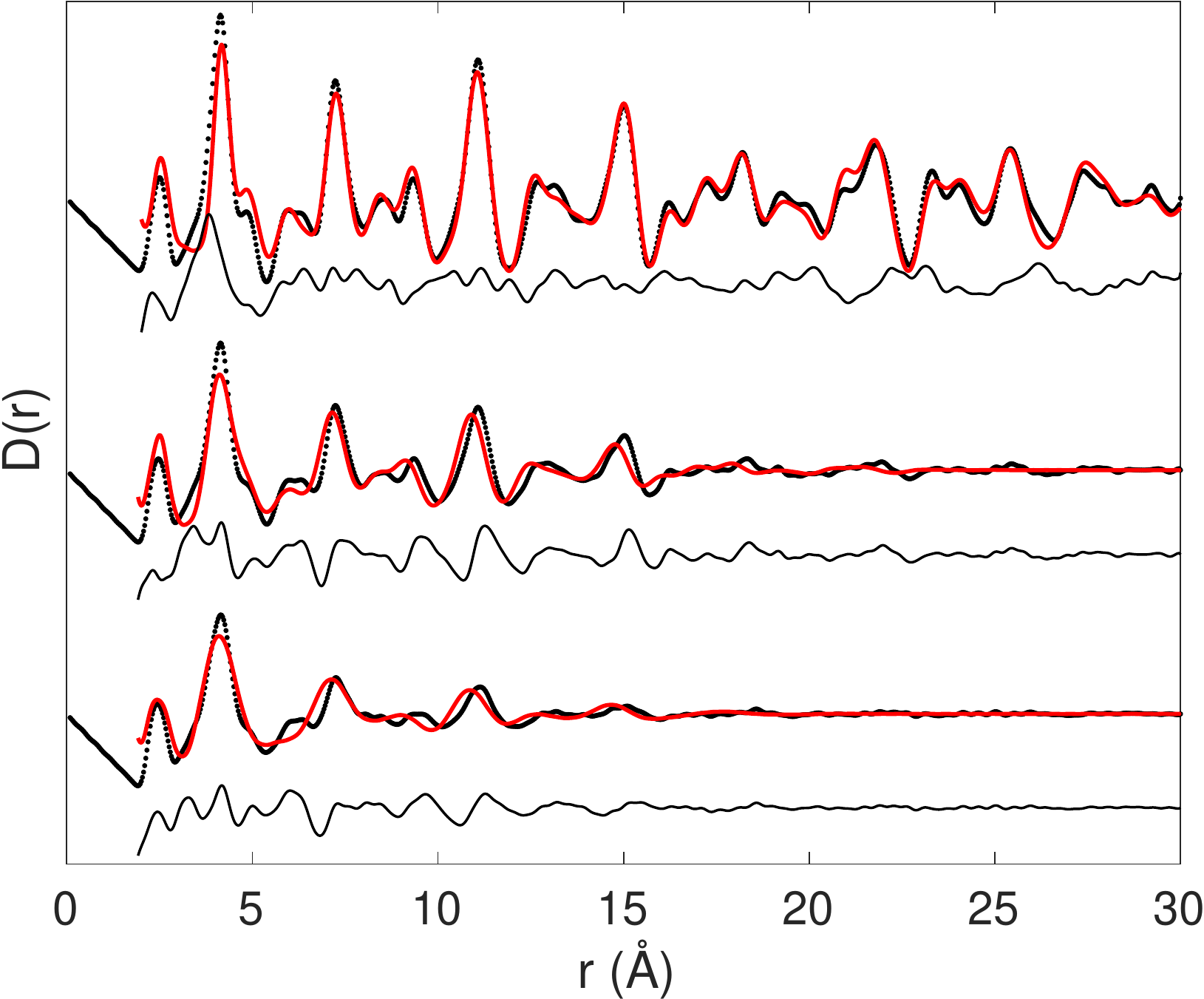}
\includegraphics[width=8cm]{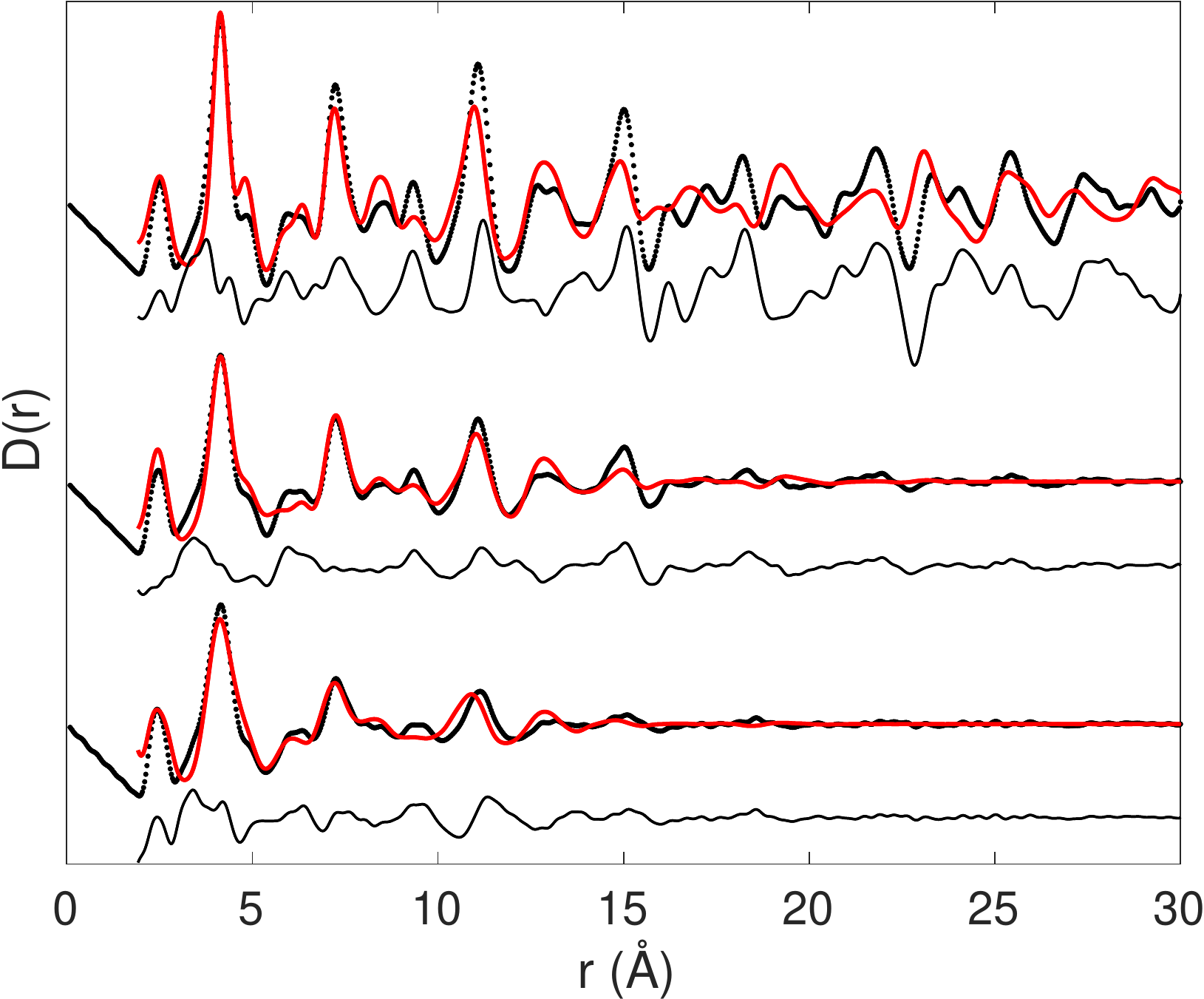}
\caption[]{The experimental blue solid dots and the calculated PDF from the refined structural model red solid line. PDF data are fitted using (left) the zinc-blende structure model with space group $P6_3mc$ and (right) the wurtizite with space group $F\overline{4}3m$}
\label{fig:fit_with_pure_phase} 
\end{center}
\end{figure}
We use PDFgui to get the fit results as shown in Figure \ref{fig:fit_with_pure_phase}. 
The residual function $R_\mathrm{w}$ \cite{Masadeh2007} is used to quantify the agreement of the calculated PDF from model the PDFgui build to experimental data. We use the wurtzite model and zinc blende model separately to fit the PDF data, and we compare the  $R_\mathrm{w}$ value of both fit in Table \ref{table:rw}. We started this simulation using the CdS bulk material, and we didn't get a good fit for wurtzite structure, with a much higher value of $R_\mathrm{w}$. The zinc blende structure gives superior fits for the bulk structure.
However, for the CdS RQD-355 and CdS RQD-439, the fits of wurtzite and zinc blende are comparable, and we can't get a very good fit as shown from the difference curves in Figure \ref {fig:fit_with_pure_phase}. The $R_\mathrm{w}$ values are reported in Table \ref{table:rw}. This indicates that CdS RQDs are not a pure phase, they may have mixed phases of wurtzite and zinc blend or contain stacking faults.

\bibliography{firstpaper_lei_SI.bib}